\documentclass[a4paper,11pt,leqno]{article}

\usepackage{graphicx}
\usepackage{epsfig}

\title{Monitoring of the Membrane Potential in Proteoliposomes with Incorporated
Cytochrome-c~Oxidase Using the Fluorescent Dye Indocyanine}
\author{Yu. A. Ivashchuk-Kienbaum\thanks{E-mail address: jurij.ivastsuk@t-online.de}}
\date{}

\begin{document}
\maketitle

\bibliographystyle{unsrt}

\begin{minipage}{\textwidth}
\begin{center}
\scriptsize Department of Chemistry/Biochemistry
Philipps-Universit{\"a}t Marburg, Germany
\end{center}
\end{minipage}

\begin{abstract}
\normalsize A method has been developed to monitor changes of the
membrane potential across vesicle membranes in real time. Using
the potential-sensitive fluorescent dye indocyanine and on the
basis of water/lipid redistribution model, a calculation procedure
has been introduced to estimate the membrane potential in vesicles
with incorporated cytochrome-c oxidase. Physical parameters, such
as vesicle size distribution and density of the lipid bilayer were
estimated and used as calculation parameters. By extrapolation of
transient potential change to zero time, the initial rate of the
potential change $(dU/dt)$ has been calculated. It is also shown,
that the initial potential change $(dU/dt)$ may be used to study
the proton/electron stoichiometry of cytochrome-c oxidase
incorporated in the vesicles.
\end{abstract}

\begin{section}{Introduction}
Cytochrome-c oxidase is the terminal enzyme of the respiratory
chain which catalyzes the four-electron reduction of molecular
oxygen to water and couples this reaction with the uptake of
protons from the mitochondrial matrix site. In this way, the free
energy of oxygen reduction is stored in form of a membrane
potential~\cite{mitch:1966}. Since the proposal of Wikstr{\"o}m, we
know that cytochrome-c oxidase, in addition to the uptake of
protons for water formation, is also capable of vectorial proton
translocation and acts as an electrogenic proton
pump~\cite{wikst:1977}. However, the molecular mechanisms and the
$H^{+}/e^{-}$-stoichiometry of proton translocation by
cytochrome-c oxidase are still unknown. In earlier studies of the
electrogenic activity of cytochrome-c oxidase, the changes of the
proton concentration across the lipid membrane have been directly
measured with a pH-microelectrode or a pH-indicator dye and
related to the protein
activity~\cite{casey:1979,oshea:1984,papa:1991,cap:1991}. These
measurements, however, are always accompanied with an uncertainty
concerning the buffer capacity of the vesicle interface which
should be taken into account when the real quantity of ejected
protons have to be calculated~\cite{prot:1983,heb:1994}. An
alternative possibility of investigating the proton translocation
is the measurement of the membrane potential developed during the
protein action. Methods to measure the membrane potential
generated by cytochrome-c oxidase have been developed earlier
based on indirect detection of the redistribution of lipophilic
cations by ion selective
electrodes~\cite{shin:1978,brown:1985,stev:1991}. Because of slow
response time of the ion-selective electrodes, the experimental
setup presented in these works could be used under steady state
conditions, but it does not allow kinetic measurements of the
membrane potential. The above considerations show that the
investigation of electrogenic activity of the cytochrome-c oxidase
and its ability of vectorial proton translocation requires an
appropriate method with a fast response time, which could allow
the recording of proton translocation events in real time.

The application of voltage-sensitive fluorescent dyes allows the
estimation of the membrane potential of cell organelles and
membrane vesicles~\cite{bash:1978,apel:1985,apel:1987}. The
mechanism of fluorescence response of the dye to an electrical
potential difference across the lipid bilayer is still a matter of
discussion~\cite{sims:1974,beeler:1981,lauger:1991}. Under certain
experimental conditions, the fluorescence alteration could be
described by a simple redistribution model~\cite{apel:1987}. Fast
response times of voltage-dependent fluorescent dyes makes them
useful in investigation of the electrogenic activity of ion pumps
reconstituted in vesicles~\cite{wagoner:1979,beeler:1981}.

In this work, I present a method for real-time-monitoring of the
membrane potential generated by the cytochrome-c oxidase using the
potential sensitive fluorescent dye indocyanine.
\end{section}

\begin{section}{Materials and Methods}

\begin{subsection}{Materials}
Asolectin, L-$\alpha$-phosphatidylcholine (type II from soybean)
and cytochrome c (type VI from horse hart) were obtained from
Sigma. Before use asolectin was purified by the method of Kagawa
and Racker~\cite{kagawa:1971}. Valinomycin was from Boehringer,
Mannheim. Indocyanine~\footnote{Indocianine:
1,3,3,1',3',3'-hexamethyl-2,2'-indodicarbocyanine iodide} (NK 529)
was purchased from Nippon Kankoh Shikiso Kenkyusho, Okayama,
Japan. The dialysis tube (pore-radius 2.4 nm) was purchased from
Serva, Heidelberg. Fluorescence experiments were carried out on a
Perkin-Elmer 650-40 fluorescence spectrophotometer.
\end{subsection}

\begin{subsection}{Buffers}
If not otherwise indicated, the buffer A contained (in mM): 10
K-Hepes, pH 7.4, 1.7 NaCl and 200 KCl. Buffer B contained 10
K-Hepes, pH 7.4, 200 NaCl and 1.7 KCl. Buffer C contained 10
K-Hepes, pH 7.4, 10 NaCl and 140 KCl.
\end{subsection}

\begin{subsection}{Isolation of Cytochrome-c Oxidase}
Bovine heart cytochrome-c oxidase was prepared from isolated
mitochondria as described in~\cite{errede:1978}.
\end{subsection}

\begin{subsection}{Determination of the cytochrome $aa_{3}$ content}
The cytochrome $aa_{3}$ content was calculated according
to~\cite{jagow:1972}. The extinction coefficients for cytochrome
$aa_{3}$ used were $\Delta$A$^{red-ox}$$_{605-630 nm}$ = 24.0
mM$^{-1}$ cm$^{-1}$, $\Delta$A$^{ox}$$_{603-650 nm}$ = 40.0
mM$^{-1}$ cm$^{-1}$, $\Delta$A$^{red}$$_{443-490 nm}$ = 204.0
mM$^{-1}$ cm$^{-1}$, $\Delta$A$^{ox}$$_{421-490 nm}$ = 140.0
mM$^{-1}$ cm$^{-1}$.
\end{subsection}

\begin{subsection}{Measurement of Oxygen Consumption}
Oxygen consumption was measured polarographically by using a
Clark-type oxygen electrode attached to a thermostatically
controlled cell~\cite{braut:1978}. 10 - 20~$\mu$l vesicles with
reconstituted cytochrome-c oxidase were introduced in 1.5 ml of
buffer A with 25 mM potassium ascorbate and various concentrations
of cytochrome c (1 - 50 $\mu$M). For measurements of the enzyme
activity 0.05\% laurylmaltosid and 0.1 mM EDTA were added
additionally. If not otherwise indicated, the experiments were
carried out at 20$^{0}$ C.
\end{subsection}

\begin{subsection}{Vesicle Preparation}
The enzyme was reconstituted into vesicles by the following
procedure: purified asolectin (30 mg/ml) and 1.5\% (w/w)
Na-cholate was sonicated to clarity in buffer C. Cytochrome-c
oxidase was added to the solubilized lipid to give a lipid to
protein ratio of 50/1 (w/w). After brief mixing, the combined
solution was transferred to 6 mm dialysis tubing and dialysed for
48~h at 4$^{0}$C against 500-fold excess volume of buffer C.
During the dialysis, buffer C was replaced twice. To calibrate the
membrane potential in control experiments, part of the vesicles
was prepared in buffer A and then dialysed additionally for 24 h
against buffer B to reach equal osmotic conditions but different
inside/outside concentrations of K$^{+}$. Protein-free vesicles
were prepared from purified asolectin by the same procedure
without the addition of protein.
\end{subsection}

\begin{subsection}{Determination of the Lipid Content}
The lipid concentration in the suspension was estimated by
determination of choline containing lipids using the phospholipid
B test~\cite{takeyama:1977}. The amount of choline containing
lipids was related to that of all lipids in asolectine
composition. The concentration of choline containing lipids in the
asolectin estimated in our laboratory (30\%) was in a good
agreement with published data~\cite{kagawa:1971}.
\end{subsection}

\begin{subsection}{Electron Microscopy of the Vesicles}
The electron microscopic imaging of the vesicles was performed
using the method of negative staining with ammonium molybdate
(NH$_{4}$)$_{6}$Mo$_{7}$O$_{24}$~\cite{haschma:1972}.
\end{subsection}

\begin{subsection}{Fluorescence Measurements}

Fluorescence experiments were carried out in a thermostatically
controlled cuvette holder which was equipped with a magnetic
stirrer. The excitation wavelength was set to 620 nm (slit width
20 nm) and the emission wavelength to 660 nm (slit width 5 nm).
The indocyanine stock solution contained 5.76 mM dye in ethanol.
From this solution dilutions were prepared daily by mixing with
ethanol. Concentrations were chosen such that addition of 1 $\mu$l
dye solution to the cuvette resulted in the desired final
concentration (1 - 3 $\mu$M). All fluorescence data were
normalised with respect to a fluorescence standard.

The cuvette was filled with 1 ml buffer and equilibrated in the
cuvette holder to the desired temperature, then 1 $\mu$l of an
indocyanine solution was added. After the fluorescence signal was
constant, an aliquot of the vesicle suspension was added.
Fluorescence changes, $\Delta F$, caused by additions of reagents
were determined as relative signal changes with respect to
fluorescence level, $F$, prior to the addition; they were
corrected for the small dilution effect which was determined
separately by adding a known amount of buffer solution.
Independent experiments (data not shown) revealed that addition of
1 - 3 $\mu$l of ethanol had no effect on the fluorescence. At the
beginning and at the end of each experiment the temperature in the
cuvette was controlled. All experiments were carried out at
20$^{0}$C.

\end{subsection}
\end{section}

\begin{section}{Results}

\begin{subsection}{Physical Parameters of the Vesicles}

The negative staining electron microscopy of the prepared vesicles
revealed a rather narrow size distribution of mainly spherical
vesicles. On the images of the vesicles an insignificant amount of
nonspherical particles could be seen which were attributed to
nonvesicular aggregates (\emph{micrographs not shown}). The size
distribution of the vesicle population was analyzed by measuring
the diameter of each vesicle followed by fitting of the data to
the normal Gaussian distribution. For the statistical analysis
about 1500 vesicle diameters were determined. By this method the
nonvesicular fragments could be excluded from the analysis.
Fig.~1$a$~and~$b$ represents the probability function of the
vesicle diameters in the range of 20 -- 200 nm.

\begin{figure}

\begin{minipage}[t]{0.5\textwidth}

\resizebox{\textwidth}{!}{\includegraphics*[20mm,30mm][200mm,300mm]{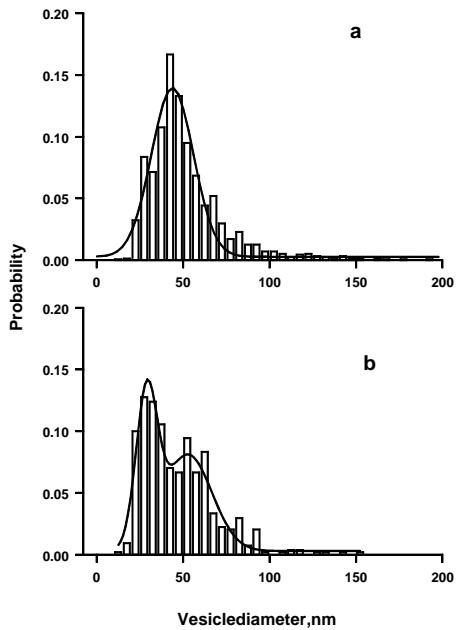}}
\end{minipage}\hfill \parbox[b]{0.4\textwidth}{\caption{\scriptsize Statistical analysis of the size distribution
of asolectin vesicles without ($a$) and with cytochrome-c oxi-dase
($b$). The analysis of the size distribution has been done by
measuring the diameter of each vesicle, followed by fitting of the
data to the normal Gaussian distribution. The estimated average
diameter for protein-free vesicles corresponds to 44 $\pm$ 12 nm,
for protein containing vesicles 52 $\pm$ 14 nm.}}

\end{figure}

The size distribution of vesicles without cytochrome-c oxidase,
Fig.1$a$, was in a good agreement with Gaussian distribution, and
the determined average diameter was 44 $\pm$ 12 nm. The size
distribution of vesicles with incorporated cytochrome-c oxidase
(Fig.1$b$) clearly shows two vesicle populations, one of
29~$\pm$~6~nm and the other of 52~$\pm$~14~nm average diameter.
The formation of two vesicle populations after protein
reconstitution could be explained if we assume that not all
vesicles in the population contain protein. This assumption has
been supported experimentally by use of ion-exchange
chromatography. The difference in diameter of protein-free and
protein-containing populations could have resulted from the
enlargement of the vesicles due to protein
incorporation~\cite{madden:1984}. The formation of protein-free
vesicles with an average diameter smaller than 44 nm is difficult
to explain so far. One explanation could be that introducing the
third component --- protein to the lipid-detergent mixture could
also influence the formation and size of protein-free
vesicles.This subdivision of two vesicle populations could be
analyzed statistically if we assume that the number of
incorporated proteins is controlled by Poisson statistics. If we
further assume that during the procedure of vesicle preparation
most of the protein is incorporated into the vesicles, the
protein--to--lipid ratio used in this work (\emph{see} vesicle
preparation) predicts an average monomeric enzyme complex per
vesicle of 1.9--2.2. According to the Poisson distribution of the
enzyme, 40\% of the vesicles should be protein free. This estimate
is in a good agreement with the distribution represented in
Fig.1$b$. In our calculations we used the molecular weight of the
monomeric complex of cytochrome-c oxidase of 204000
Daltons~\cite{kaden:1986}. It is still not clear, however, whether
the monomer or dimer is the functional unit of the reconstituted
cytochrome-c oxidase.

To study the distribution of indocyanine between the water- and
lipid phases, we need to know the volume of asolectine in the
vesicle probe. This could be calculated if we know the density of
the asolectine used. The size of the vesicle and the number of
lipid molecules assembled in a vesicle give us the information
about the average molecular weight and density of the asolectine.
These calculations are based on data published
earlier~\cite{huang:1978}. From these data we assume that the
radius of the head group is 4.85 \AA{} in the outer and 4.41 \AA{}
in the inner monolayer. Therefore, the area occupied by a lipid
molecule in both monolayers can be calculated to be 74~\AA$^{2}$
and 61~\AA$^{2}$, respectively. Protein-containing vesicles have
an external diameter of 52~$\pm$~14~nm (\emph{see above}). If the
lipid bilayer is assumed to be packed with the highest possible
density of lipid head groups, a vesicle contains
8.273~$\cdot$~10$^{4}$ lipid molecules with a deviation between
4.292~$\cdot$~10$^{4}$ and 1.355~$\cdot$~10$^{5}$ molecules due to
the standard deviation of the vesicle diameter.

To calculate the average molecular weight and density of the
asolectin we have measured the entrapped volume for a known
quantity of lipid. Vesicles were prepared as described above with
known concentration of K$_{2}$CrO$_{4}$, separated on a
Sephadex~G25 column and the lipid concentration was estimated. The
whole entrapped volume was determined spectroscopically, assuming
that the K$_{2}$CrO$_{4}$ concentration inside the vesicles
remains constant. The average molecular weight of asolectin
estimated in such a way were 880 g/mol and 1028 mg/ml
respectively.

\end{subsection}

\begin{subsection}{Spectral Properties of Indocyanine}

Indocyanine has its fluorescence emission maximum at 658~nm, and
its absorption maximum at 635.7 nm. The additional shoulders
between 570--600 nm (absorption spectra) and 690--720 nm
(fluorescence spectra) indicate the presence of higher aggregates
of the dye in aqueous solution, as has been suggested
by~\cite{sims:1974}. The fluorescence maximum of cyanine dyes
shifts further to the red when one transfers the dye from water
into nonpolar solvents. Consistent with this finding, the
fluorescence maximum of indocyanine is shifted from 658~nm to
676~nm. This is shown in Fig. 2, which represents the fluorescence
spectra of indocyanine as a function of the concentration of
protein-free vesicles. Not only a red shift of the fluorescence
spectra, but also a quenching effect could be seen during the
vesicles addition, which is well-defined in the lipid
concentration range between 0 and 60 $\mu$g/ml.

\begin{figure}
\begin{minipage}{0.5\textwidth}

\rotatebox{-90}{\resizebox{\textwidth}{!}{\includegraphics*[20mm,40mm][195mm,270mm]{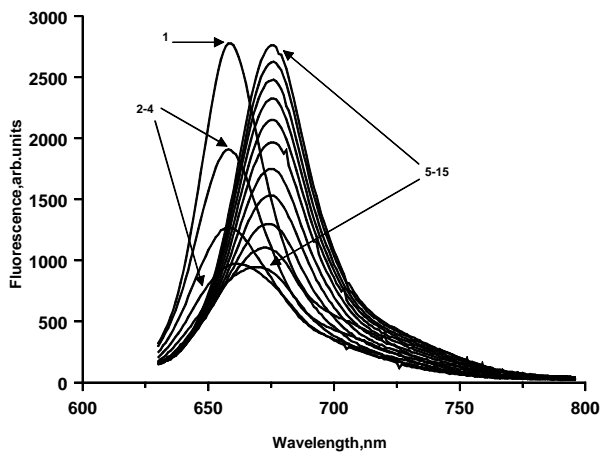}}}
\end{minipage}\hfill \parbox{0.3\textwidth}{\caption{\scriptsize Dependence of the fluorescence of indocyanine
from the quantity of the vesicles in the water solution. 1,
fluorescence spectra of the indocyanine (2.88 $\mu$M) in buffer A
(in mM: 140 KCl, 10 NaCl, 10 Hepes, pH~=~7.4). 2--15, consecutive
addition of aliquots (15 $\mu$g asolectin) of the protein-free
vesicle suspension. The excitation wavelength was 620~nm, slit
width 20~nm and 5~nm.}}

\end{figure}

A fluorescence change similar to that seen after vesicle addition
could be observed when a membrane potential (inside negative)
across the lipid bilayer of the vesicle was generated. The exact
mechanism which is involved in the fluorescence change during the
polarization of the vesicle membrane is still unknown. It has
become evident, however, that the behavior of many potential
sensitive fluorescent dyes can be described by a distribution
model~\cite{sims:1974,apel:1985,apel:1987,zouni:1993}. Normally
the lipophilic cation indocyanine is distributed between the water
and lipid phases according to its partition coefficient.
Polarisation of the lipid membrane by the generated membrane
potential leads to redistribution of the dye molecules between the
water and lipid phases because of electrostatic attraction between
the positively charged indocyanine and negatively charged vesicle
interior.

As can be seen from Fig. 2, there were two fluorescence effects
which appeared during the vesicle addition. First, there was a
general red shift of the spectra of indocyanine (658 nm
$\rightarrow$ 676 nm), and second, there was a pronounced
quenching effect at the lipid concentrations 0--60 $\mu$g/ml
(additions 2--4, Fig. 2). If we consider the distribution model,
with an increasing vesicle concentration the amount of
membrane-bond dye molecules also increases, and consequently the
concentration of the dye molecules in water decreases. This would
lead to a pure shift-effect of the indocyanine spectra with a
constant quantum yield. The position of the peak maximum would
then reflect the dye-bound/dye-free ratio. In fact, we observe not
only the red shift of the peak maximum, but additionally there is
a strong quenching effect which makes the analysis of the dye
distribution more complex.  It seems that the most appropriate
description of the dye-vesicle interaction is a mechanism of dye
binding to a vesicle with a discrete number of binding
sites~\cite{clarke:1991,clarke:1992} If we assume that the initial
concentration of dye in the water phase is sufficient for the
saturating occupation of binding sites at small vesicle
concentration and, because of that, the quenching of the
fluorescence of the bound dye molecules occurs due to their
aggregation, the fluorescence change in this case would be
determined mainly by the change of the dye concentration in water.
This is what we observe in Fig. 2. After the vesicle addition
(2-4) the fluorescence intensity decreased and the position of the
peak (658nm) remained nearly unchanged. If after the addition of
small amounts of vesicles to the appropriate concentration of dye
not all of the binding sites are occupied, it would still be
possible to apply the distribution model to describe the partition
of the dye between water and lipid vesicles. The partition
coefficient will then express the affinity of binding dye
molecules to the vesicles~\cite{zouni:1993}.

In Fig. 3 the fluorescence change of indocyanine is shown as a
function of the lipid concentration at various fixed wavelengths.
The wavelength at which linearity combines with a large
fluorescence change in response to the change in vesicle
concentration ($dF/dC$ lipid) would be the appropriate fixed
wavelength to monitor the dye-vesicle interaction. In the
following experiments a fixed wavelength of 660 nm was chosen
because, in the range of lipid concentrations 5 - 30 $\mu$g/ml,
the negative fluorescence change shows approximate linearity and
the magnitude of $dF/dC$ lipid is comparatively high.

\begin{figure}
\begin{minipage}{0.55\textwidth}

\rotatebox{-90}{\resizebox{\textwidth}{!}{\includegraphics*[20mm,30mm][195mm,280mm]{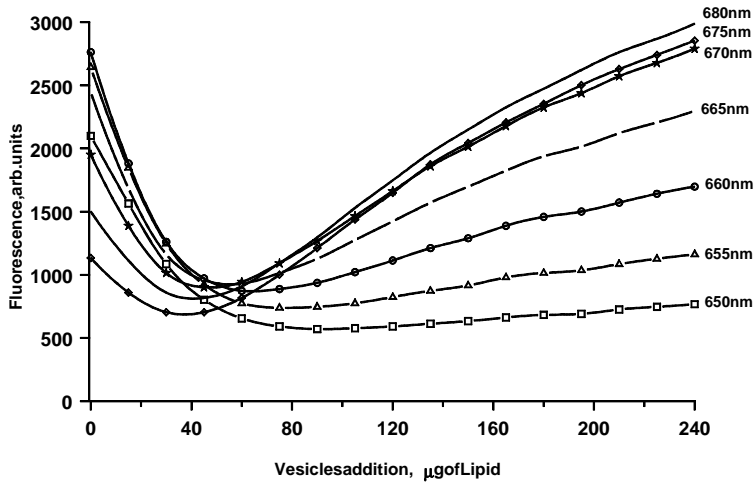}}}
\end{minipage}

\begin{minipage}{\textwidth}
\caption{\scriptsize Dependence of the fluorescence of indocyanine
from the quantity of vesicles in the solution at various fixed
emission wavelengths. Initial concentration of indocyanine was
2.88 $\mu$M in buffer A (140 mM KCl, 10 mM NaCl, 10 mM Hepes,
pH~=~7.4).}
\end{minipage}
\end{figure}

In the absence of lipid vesicles, the total fluorescence ($F_{0}$)
is determined only by the amount of dissolved dye molecules. The
fluorescence of indocyanine in water is a function of its
concentration, ionic strength (and/or buffer composition) and
aggregation state of the dye molecules. In Fig. 4 the fluorescence
of indocyanine ($F_{0}$) at 660 nm can be seen as a function of
its aqueous concentration. $F_{0}$ is referred to the fluorescence
intensity measured with a fluorescence standard and expressed in
dimensionless numbers. To describe this dependence we have used
the following equation:

\begin{equation}
F_{0} = Fm\left[1-exp\left(-\frac{c_{w}}{\beta}\right)\right]
\end{equation}

where $Fm$ is the maximum fluorescence, $c_{w}$ is the
concentration of indocyanine in water and $\beta$ is a constant
parameter which represents a value proportional to the magnitude
of the inner filter effect of indocyanine. Because of the
significant degree of overlap between the absorbance and
fluorescence spectra of indocyanine (absorption max. at 636,
fluorescence max. at 658), its fluorescence spectra is not linear.
The equation (1) represents the solution of the differential
equation (2):
\begin{equation}
\frac{dF_{0}}{dc_{w}} = Fm - \frac{F_{0}}{\beta}
\end{equation}

As can be seen in Fig. 4, equation (1) describes quite well the
concentration dependence of the fluorescence of indocyanine in the
range of 0 - 4 $\mu$M.

\begin{figure}
\begin{minipage}[b]{0.5\textwidth}

\rotatebox{0}{\resizebox{\textwidth}{!}{\includegraphics*[20mm,15mm][199mm,280mm]{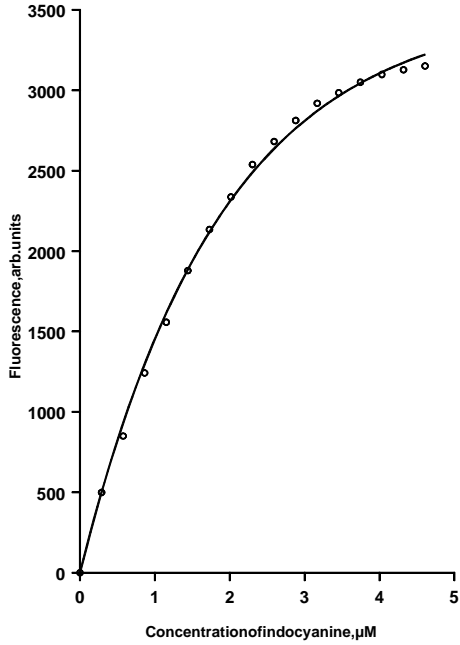}}}
\end{minipage}\hfill \parbox[b]{0.35\textwidth}{\caption{\scriptsize Dependence
of the fluorescence of indocyanine on its concentration in
solution. Measurement buffer was: 140 mM KCl, 10 mM NaCl, 10 mM
Hepes, pH = 7.4. Open circles are measured values; the smooth line
is the fitted function eqn. (1) with parameters: $Fm$ = 3527,
$\beta$~=~1.883}}

\end{figure}
\end{subsection}

\begin{subsection}{Estimation of the partition coefficient at zero membrane potential}
The membrane-water partition coefficient, $\gamma$ , is defined
by:
\begin{equation}
\gamma = \frac{c_{l}}{c_{w}} =
\frac{\frac{n_{l}}{V_{l}}}{\frac{n_{w}}{V_{w}}}
\end{equation}
where $c_{l}$ and $c_{w}$ are the concentrations of the dye in the
lipid phase and in the water phase, $n_{l}$ and $n_{w}$ are the
amounts of dye in the lipid phase and in the water phase and
$V_{l}$ and $V_{w}$ are the volumes of the lipid and water phases.
If $n_{0}$ is the total amount of the dye, the amount of the dye
located in the lipid bilayer can be calculated by:
\begin{equation}
n_{l} = n_{0} - n_{w}\quad\mbox{or}\quad n_{l} = V_{w}\left(c_{0}
- c_{w}\right)
\end{equation}
Where $c_{0}$ and $c_{w}$ are the initial and postredistribution
concentrations of dye in the water phase. Measuring the dye
concentration before and after addition of vesicles we can
estimate the membrane-water partition coefficient of indocyanine,
which is defined by:
\begin{equation}
\gamma = \frac{c_{l}}{c_{w}} =
\frac{\left(c_{0}-c_{w}\right)V_{w}}{c_{w}V_{l}}
\end{equation}
Rearranging eqn. (1) gives the following expressions for $c_{0}$
and $c_{w}$:
\begin{equation}
c_{0} = -\beta \ln\left(1-\frac{F_{0}}{F_{m}}\right)
\end{equation}
\begin{equation}
c_{w} = -\beta \ln\left(1-\frac{F}{F_{m}}\right)
\end{equation}
and the partition coefficient will be defined as:
\begin{eqnarray}
\gamma = \frac{\left[\beta \ln\left(1 - \frac{F}{F_{m}}\right) -
\beta \ln\left(1 -
\frac{F_{0}}{F_{m}}\right)\right]V_{w}}{\left[-\beta \ln\left(1 -
\frac{F}{F_{m}}\right)\right]V_{l}} =\nonumber\\ \frac{\ln
\left(F_{m} - F\right) - \ln \left(F_{m} -F_{0}\right)}{\ln F_{m}
- \ln \left( F_{m} - F\right)} \cdot \frac{V_{w}}{V_{l}}
\end{eqnarray}
where $F_{0}$ is the fluorescence of indocyanine in water in
absence of lipid vesicles, $F$ -- the fluorescence of the dye
after addition of a given amount of vesicles and $F_{m}$ was
obtained as a fitted parameter from Fig. 4.

Fig. 5 represents the estimation of the partition coefficient at 0
mV for various vesicle concentrations in the solution. The volume
of the lipid phase was calculated from the density of asolectin
(1.0276 $\cdot$ $10^{3}$ mg/ml), \emph{see above}, and the volume
of the water phase ($V_{w}$) was 1 ml. The slope of the straight
line in Fig. 5 defines the partition coefficient which corresponds
to $\gamma$~=~68.5624~$\cdot$~$10^{3}$.

\begin{figure}
\begin{minipage}{0.6\textwidth}

\rotatebox{-90}{\resizebox{\textwidth}{!}{\includegraphics*[15mm,15mm][199mm,280mm]{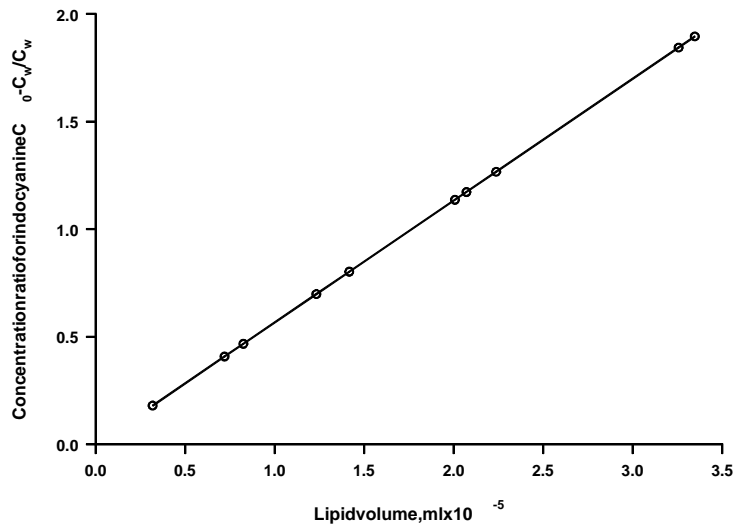}}}
\end{minipage}

\begin{minipage}{\textwidth}{\caption{\scriptsize Estimation of the partition coefficient at 0
mV voltage. $C_{0}$ - is the initial concentration of indocyanine
in the water phase, $C_{w}$ - is the concentration of indocyanine
in the water phase after redistribution. The slope corresponds to
the partition coefficient at 0 mV voltage,
$\gamma$~=~68.5624~$\cdot$~$10^{3}$. The lipid volume was
calculated from the estimated asolectin density 1028~mg/ml
(\emph{see text}), water volume ($V_{w}$) was 1 ml.}}
\end{minipage}

\end{figure}
\end{subsection}

\begin{subsection}{Calibration of the fluorescence signal as a function of membrane potential}
Membrane potentials may be generated by establishing a
transmembrane K$^{+}$ concentration difference in the presence of
valinomycin. At the beginning of the experiment the vesicle
interior contains 200 mM KCl (buffer A); the extravesicular
aqueous space contains 1.7 mM KCl (buffer B). The addition of
valinomycin to the solution makes the vesicle membrane selectively
permeable for the K$^{+}$ cations. The intravesicular
concentration of K$^{+}$ is decreased by the release of K$^{+}$
ions, which builds up a Nernst potential of 120.4 mV (inside
negative), according to:
\begin{equation}
U = \Psi^{'} - \Psi^{''} = \frac{RT}{F} \cdot \ln
\frac{c^{''}}{c^{'}}
\end{equation}
$\Psi^{'}$ and $\Psi^{''}$ are the intra- and extravesicular
potentials, respectively, and $c^{'}$ and $c^{''}$ are the
corresponding K$^{+}$ concentrations; $R$ is the gas constant, $T$
the absolute temperature and $F$ the Faraday constant.

After the inside-negative membrane potential is generated, further
accumulation of indocyanine in the lipid phase occurs. According
to the new equilibrium state, the partition coefficient changes.
The new (apparent) partition coefficient ($\gamma_{app}$) of
indocyanine is related to the generated membrane potential and can
be used for calculation of the membrane potential
value~\cite{apel:1987}.

The accumulation of the dye in the lipid bilayer leads to a
fluorescence decrease. The potential-induced fluorescence change
was expressed by the relative fluorescence change, Fig. 6:
\begin{equation}
\frac{\Delta F}{F} \equiv \frac{F - F_{u}}{F}
\end{equation}
$F_{u}$ is the fluorescence estimated after the membrane potential
generation.

\begin{figure}
\begin{minipage}{0.6\textwidth}

\rotatebox{-90}{\resizebox{\textwidth}{!}{\includegraphics*[15mm,15mm][199mm,280mm]{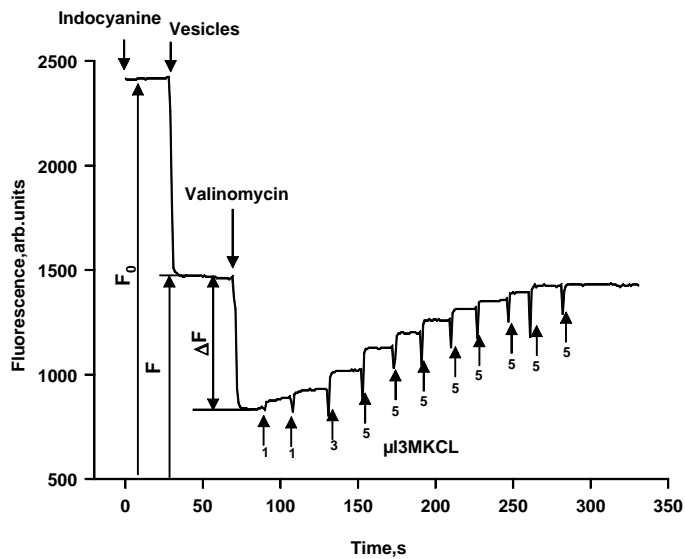}}}
\end{minipage}

\begin{minipage}{\textwidth}{\caption{\scriptsize Calibration experiment.
Relationship between the fluorescence and the defined membrane
potential. $F$ is the fluorescence intensity in arbitrary units.
At the beginning of the experiment 1 ml of buffer (in mM: 10
K-Hepes, pH 7.4, 140 NaCl and 10 KCl) was present in the
fluorescence cell. Thereafter the following additions were made:
1~$\mu$l indocyanine solution (final concentration 2.88~$\mu$M)
($F_{0}$ is the initial fluorescence intensity in the cell);
1~$\mu$l (30~mg/ml lipid) vesicles with 140~mM K$^{+}$, 10~mM
Na$^{+}$, 10~mM K-Hepes and pH~7.4 inside ($F$ is the fluorescence
intensity at 0 mV); 1~$\mu$l valinomycin (final concentration
0.02~$\mu$M), $\Delta$F -- is the fluorescence change due to the
membrane potential estimated (66,63 mV); consecutive additions of
the 3~M KCl solution to change the defined voltage value.}}
\end{minipage}
\end{figure}

Estimation of the apparent partition coefficient, $\gamma_{app}$,
at several known membrane potentials and its relation to the
membrane-potential dependent fluorescence change,  F/F, revealed
the functional dependence between membrane potential and the
relative fluorescence change. The apparent partition coefficient
could be defined analogously to the partition coefficient at zero
voltage as:
\begin{equation}
\gamma_{app} = \frac{\ln \left( F_{m} - F_{app} \right) - \ln
\left( F_{m} - F_{0} \right)}{\ln F_{m} - \ln \left (F_{m} -
F_{app} \right)} \cdot \frac{V_{w}}{V_{l}}
\end{equation}
where $F_{app}$ is the fluorescence intensity at a predefined
voltage value. The calibration of $\gamma_{app}$ was carried out
as shown in Fig. 6. Addition of indocyanine to a final
concentration of 2,88 $\mu$M resulted in the initial fluorescence
level ($F_{0}$). After that, the vesicle suspension (usually 1 - 3
$\mu$l) was added. The fluorescence decreased due to the dye
distribution at zero voltage (in the absence of valinomycin) and
reached the ($F$) level. Because of different K$^{+}$
concentrations in the intra- and extravesicular medium, upon the
addition of valinomycin a membrane potential was generated
according to the Nernst equation (9). When the K$^{+}$
concentration gradient was diminished by addition of small volumes
(1 - 5 $\mu$l) of 3 M KCl solution, the membrane potential
decreased and the fluorescence intensity increased. In this way
the relative change of fluorescence intensity ($\Delta$$F$/$F$)
and the apparent partition coefficient ($\gamma_{app}$) were
determined for different fixed membrane potentials using equations
(10) and (11). The values of partition coefficients were
determined separately for vesicle preparations with and without
protein as shown in Fig. 7. The calculated curve was obtained by
fitting the experimental data to the polynomial equation:
\begin{equation}
\gamma_{app} = \gamma + A_{1} \cdot U + A_{2} \cdot U^{2}
\end{equation}
where $\gamma$ is the partition coefficient at 0 voltage, $U$ is
the membrane potential value, and $A_{1}$ and $A_{2}$ are fitting
parameters ($A_{1}$ = 1162.7 mV$^{-1}$, $A_{2}$ = 11.29
mV$^{-2}$). The equations (11) and (12) allowed the calculation of
the relative fluorescence change for any desired voltage and
vesicle concentration in the range of 5 -- 30 $\mu$g/ml lipid
concentration. The relation between relative fluorescence change
($\Delta F/F$) and appropriate $\Delta U$ could be found by the
interpolation of calculated voltage values.

\begin{figure}
\begin{minipage}{0.5\textwidth}

\rotatebox{-90}{\resizebox{\textwidth}{!}{\includegraphics*[15mm,15mm][199mm,280mm]{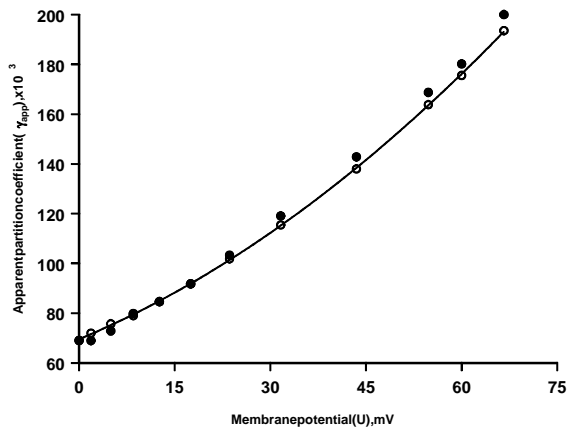}}}
\end{minipage}

\begin{minipage}{\textwidth}{\caption{\scriptsize Apparent partition coefficient,
$\gamma_{app}$, as a function of the membrane potential. Open
circles are protein free vesicles, solid circles are protein
containing vesicles. The theoretical curve was obtained by fitting
the experimental data to the polynomial equation (12), where
$\gamma$ is the partition coefficient at 0 voltage $\gamma$ =
68.5624 $\cdot$ $10^{3}$, $U$ is the membrane potential value,
$A_{1}$ and $A_{2}$ are the fitting parameters ($A_{1}$ = 1162.7
mV$^{-1}$, $A_{2}$ = 11.29 mV$^{-2}$).}}
\end{minipage}
\end{figure}

\end{subsection}
\begin{subsection}{Membrane potential generated by cytochrome-c oxidase}
Cytochrome-c oxidase reconstituted in lipid vesicles formed by
cholate dialysis exhibits two orientations. The right-side out
orientation (cytochrome-c binding site oriented outside) however,
dominates with 60~-~80\%~\cite{tihova:1993}. Addition of
cytochrome-c to the extravesicular medium (in the presence of
K$^{+}$-ascorbate) resulted in activation of enzyme molecules with
right-side out orientation. Since the cytochrome-c oxidase works
as a proton pump and reduces oxygen to water by the uptake of
protons from the vesicle inside, activation of right-side out
oriented  proteins generated an inside negative membrane
potential. The electrogenic effect of cytochrome-c oxidase is
represented in Fig. 8. After addition of 1-2 $\mu$l of vesicle
suspension the fluorescence decreased and became stable at a
fluorescence level which corresponded to the partition coefficient
of the indocyanine at 0 mV voltage. Addition of K$^{+}$-ascorbate
to a final concentration of 1 mM did not influence the
fluorescence. Further addition of cytochrome-c to a final
concentration of 10 - 50 $\mu$M, resulted in activation of the
cytochrome-c oxidase and generation of a membrane potential which
can be calculated from the change of the relative fluorescence (
$\Delta F/F$) with time (Fig. 9).

\begin{figure}
\begin{minipage}{0.55\textwidth}

\rotatebox{-90}{\resizebox{\textwidth}{!}{\includegraphics*[15mm,15mm][199mm,280mm]{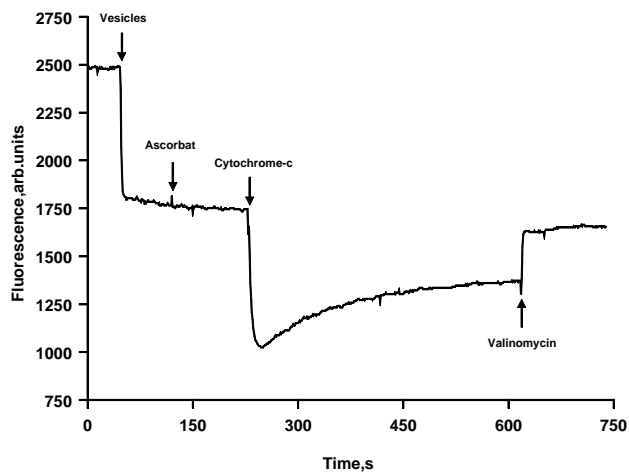}}}
\end{minipage}

\begin{minipage}{\textwidth}{\caption{\scriptsize Fluorescence change on the basis of
membrane potential development by vesicles with incorporated
cytochrome-c oxidase. Vesicles were reconstituted in buffer A (in
mM: 140 KCl, 10 NaCl, 10 Hepes, pH = 7.4). The average amount of
reconstituted proteins per vesicle was determined to be $\approx$
2. The addition of 1-2 $\mu$l of vesicle, 1 $\mu$l of 1 M
K$^{+}$-ascorbate solution to a final concentration 1 mM and 2
$\mu$l 10 mM cytochrome-c solution to a final concentration of
20~$\mu$M, results in activation of the cytochrome-c oxidase and
generation of the membrane potential. At 620 sec valinomycin was
added to decrease the membrane potential.}}
\end{minipage}
\end{figure}

\begin{figure}
\begin{minipage}{0.55\textwidth}

\rotatebox{-90}{\resizebox{\textwidth}{!}{\includegraphics*[15mm,15mm][199mm,280mm]{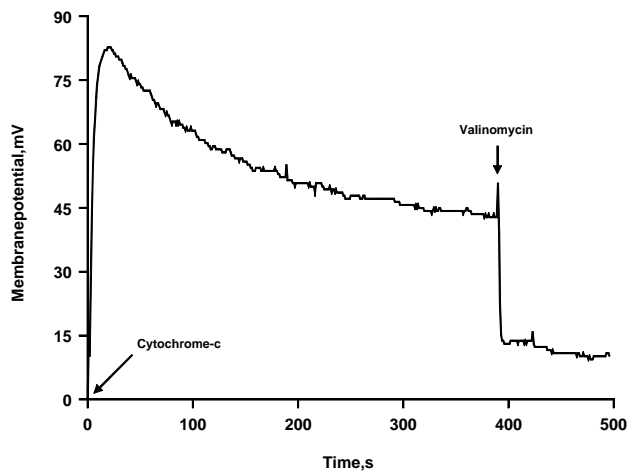}}}
\end{minipage}

\begin{minipage}{\textwidth}{\caption{\scriptsize Membrane potential development calculated
from the transient change of the relative fluorescence ($\Delta F/F$) from the experiment in Fig. 8.
Time point zero is related to the moment of addition of the cytochrome-c.}}
\end{minipage}
\end{figure}

\end{subsection}

\begin{subsection}{Influence of cytochrome-c and K$^{+}$-ascorbate addition}
In Fig. 8 it can be seen that addition of 1 $\mu$l of 1 M
K$^{+}$-ascorbate solution to the vesicle-indocyanine suspension
has no influence on the fluorescence. A final concentration
between 10 - 25 mM caused a fluorescence change ($\Delta F/F$) in
the range of only 0.5 - 1\%.

Due to its absorbance at 550 nm, addition of cytochrome-c
decreased the intensity of the fluorescence. This effect was
measured in experiments with protein free vesicles. The dependence
revealed linearity in the concentration range between 0 and 100
$\mu$M cytochrome-c, and corresponds to a fluorescence change,
$\Delta F/F$, of 0.5\% per 1 $\mu$M cytochrome-c added (data not
shown). This fluorescence change has been taken into account in
further calculations of the membrane potential by subtraction of
this fluorescence change from the registered transient data of
$\Delta F/F$, produced by the generated membrane potential. The
cytochrome-c effect was estimated by analogous measurements with
the protein-free vesicles.

\end{subsection}

\begin{subsection}{Time resolution}
One of the restrictions of this experimental set-up is the time
resolution of the initial fluorescence change upon the enzyme
activation. The time necessary to reach homogeneity after addition
of a component, or the "stirring~time", was measured as follows.
To the cuvette containing indocyanine (final concentration approx.
2.88 $\mu$M) in buffer C, 1 - 2 $\mu$l of the vesicles (without
protein) were added. After the dye redistribution equilibrium  was
reached, various portions (1 - 10 $\mu$l) of a 10 mM cytochrome-c
solution were added and the time until the fluorescence reached a
constant value was measured. Since the vesicles were protein free,
the solution of cytochrome-c produced a reduction in fluorescence
intensity caused only by absorption at 550 nm. Therefore, the
fluorescence equilibration time after the cytochrome-c addition is
the "stirring~time" of our experimental set-up. This stirring time
was estimated to be 400 - 500 ms from an average of 10
measurements. The redistribution time of the dye molecules between
water and lipid phase, however, is much shorter and lies between
100 $\mu$s and 10 ms~\cite{ross:1977}. For real time measurements
it is short enough to detect the membrane potential change even
after one enzyme turnover.

\end{subsection}

\begin{subsection}{Influence of the pH on the fluorescence of indocyanine}

The change of pH value induced by the activity of the cytochrome-c
oxidase may influence the fluorescence of indocyanine and/or its
ability to interact with the lipid bilayer. To test this
suggestion we studied the effect of the pH change on the
fluorescence. The relative fluorescence change ($\Delta F/F$)
corresponds to 4.7\% per pH unit. Due to an appropriate buffering
capacity (10 mM Hepes), a pH change during the action of
cytochrome-c oxidase could not be more than 0.001 pH units.
Therefore, the change of relative fluorescence due to pH change is
negligible and can be ignored in our calculations.

\end{subsection}

\begin{subsection}{Influence of indocyanine on the function of cytochrome-c oxidase}

\begin{figure}
\begin{minipage}{0.55\textwidth}

\rotatebox{-90}{\resizebox{\textwidth}{!}{\includegraphics*[15mm,15mm][199mm,280mm]{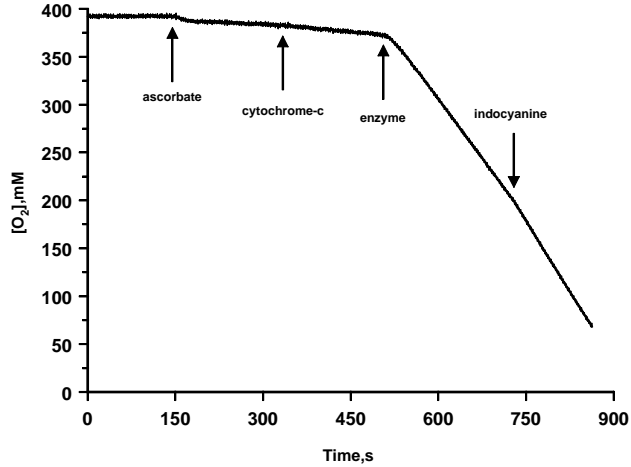}}}
\end{minipage}

\begin{minipage}{\textwidth}{\caption{\scriptsize Influence of indocyanine on the activity of
cytochrome-c oxidase. Oxygen consumption was measured as has been
described in "Materials and Methods". The enzyme activity was
measured in: 40 mM KCl, 10 mM Hepes, 0.05\% laurylmaltoside, 0.1
mM EDTA, 25 mM K$^{+}$-ascorbate and 50 $\mu$M cytochrome-c. The
turnover number (TN) is expressed in e$^{-}$ s$^{-1}$
heme~$aa_{3}^{-1}$. Before indocyanine addition the TN was 95
s$^{-1}$, after indocyanine addition the TN was 106 s$^{-1}$.}}
\end{minipage}
\end{figure}

In Fig. 10 the influence of indocyanine addition on the activity
of cytochrome-c oxidase is shown. After addition of cytochrome-c
the turnover rate measured corresponded to 96 e$^{-}$ per s and
heme $aa_{3}$. Addition of indocyanine (final concentration 2.88
$\mu$M) increases the turnover rate by 10-15\%. Normally the
standard deviation of the oxygen consumption measurements lay
between 5-10\%. Therefore, for more accurate measurements this
deviation should be taken into account but could be ignored for
the present studies.

\end{subsection}
\begin{subsection}{The initial rate of membrane potential change}
A biological membrane can be described physically by two
parameters, the specific membrane capacitance, $C_{m}$, and (leak)
membrane conductance, $\Lambda_{m}$. The change of the membrane
potential, $dU/dt$, combines these parameters and the electrogenic
activity by the following relation:
\begin{equation}
\frac{dU}{dt} = \frac{n_{p} \cdot e_{0} \cdot k_{s} \cdot \nu}{A
\cdot C_{m}} - \frac{\Lambda_{m} \cdot U}{C_{m}}
\end{equation}
where $n_{p}$ is the number of outside-out (functionally) oriented
protein molecules per vesicle, $e_{0}$ the elementary charge,
$k_{s}$ is the stoichiometry factor (translocated protons per
turnover), $\nu$ the turnover rate of the pump and $A$ the average
area of the vesicle membrane. Since immediately after enzyme
activation the membrane potential is close to 0, the leak current
term, $(\Lambda_{m}U)/C_{m}$, can be neglected and eq. 13 reduces
to \cite{apel:1987,iva:1993,iva:1994}:
\begin{equation}
\frac{dU}{dt} = K \cdot k_{s} \cdot \nu
\end{equation}
where $K$ contains all substrate independent factors of Eq. 14:
\begin{equation}
K = \frac{n_{p} \cdot e_{0}}{A \cdot C_{m}}
\end{equation}

This analysis shows that under standard conditions the initial
rate of membrane potential change depends only on the
stoichiometry factor ($k_{s}$) and the turnover rate of
cytochrome-c oxidase incorporated in the vesicle membrane ($\nu$).
Fig. 11 shows a complete time course of an experiment. For the
calculation of the initial rate of the membrane potential
generation we have used the data from Fig. 9 in the range of 0 -
25 sec, fitted to the function $U = U_{max} \cdot
\left[1-exp\left(-t/\tau \right)\right]$, and differentiated with
respect to t to obtain the initial rate (dU/dt) at time point 0.
The greatest calculated potential value reached was 79.05 mV and
the initial rate of potential change corresponded to 24.1 mV/sec.
The variation of these two parameters could be used further in the
study of proton/electron stoichiometry of cytochrome-c oxidase.
\begin{figure}
\begin{minipage}{0.6\textwidth}

\rotatebox{-90}{\resizebox{\textwidth}{!}{\includegraphics*[15mm,15mm][199mm,280mm]{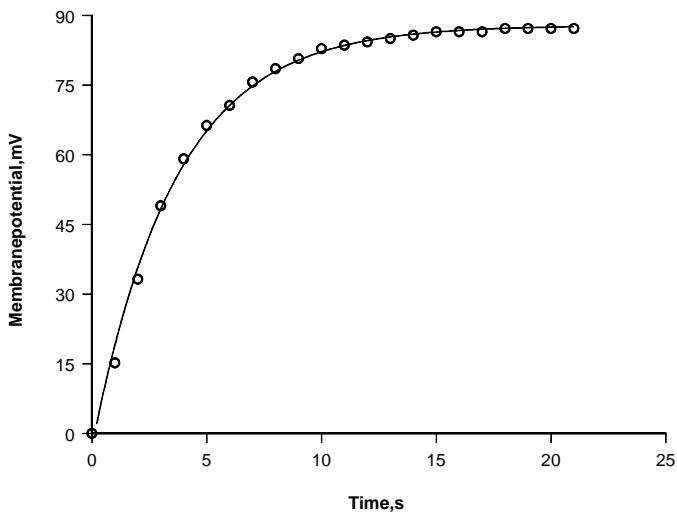}}}
\end{minipage}

\begin{minipage}{\textwidth}{\caption{\scriptsize Time resolved membrane potential change.
Time resolution was performed in the range of 0 - 25 sec from the measured
and calculated membrane potential data (Fig.9). Open circles are measured data,
smooth line are fitted data to the function
$U = U_{max}\cdot \left[1-exp\left(-t/\tau \right)\right]$,
were $t$ is time in sec, $U_{max}$ = 79.05 mV and $\tau$ = 3.28 sec
are fitting parameters. From these data the calculated $dU/dt$ corresponds to 24.1 mV/sec.}}
\end{minipage}
\end{figure}

\end{subsection}

\end{section}
\begin{section}{Discussion}
The ability of lipophilic fluorescent dyes to change their
spectroscopic properties according to the membrane potential
applied across the lipid bilayer is well
known~\cite{sims:1974,beeler:1981,wagoner:1979}. The mechanism of
voltage response of some potential-sensitive dyes, on the other
hand, is not understood so far~\cite{wagoner:1979,clarke:1992}.
But in many cases, especially for cyanine or oxonol dyes, the
response mechanism could be adequately described by a distribution
model~\cite{apel:1985,apel:1987,zouni:1993}. Depending on the
polarity of its environment the indocyanine dye displays different
fluorescence emission spectra. The fluorescence maximum shifts to
red wavelengths when the dye molecule moves from water into the
lipid phase (Fig. 2). The dye redistribution events can be
detected by measuring the fluorescence change in the range 650 nm
-- 690 nm. In this paper the use of the indocyanine dye to monitor
membrane potential in vesicles with cytochrome-c oxidase has been
shown. Although it seems that a binding model with a discrete
number of binding sites is more suitable for the description of
the dye-vesicle interaction, I show here that at high dye/vesicle
ratios the simple distribution model could be used. The
restriction of this method is the concentration of the vesicles,
which should be in the range of up to 30 $\mu$g/ml lipid at dye
concentrations of about 2.5-2.9 $\mu$M. At 660 nm we could
register a relative fluorescence change, $\Delta F/F$, of 7 \% per
10 mV of membrane potential change.

The time resolution of this experimental set-up is determined
mainly by two time constants. First, the "stirring time", which is
the time until the solution reaches its homogeneity in the cuvette
and second, the time response of the fluorescent dye. The stirring
time was the main time limiting factor and corresponded to 400 -
500 ms. The time response of the indocyanine dye is considerably
shorter. Ross et al. \cite{ross:1977} observed a biphasic
fluorescence signal from nerve axons stained with indocyanine
after application of a voltage step, consisting of a fast
component with a time constant below 100 $\mu$sec and a second
slower component with a time constant of about 10 msec. Since the
turnover number of cytochrome-c oxidase is in the range of 50 --
100 sec$^{-1}$, the response time of the indocyanine is short
enough to detect the change of the membrane potential without
deformation even after one enzyme turnover.

The time course of the relative fluorescence change after the
addition of cytochrome-c  was measured and used to calculate the
membrane potential change. Extrapolation of this voltage course to
the time point 0, gives us a function of the membrane potential
development with time. By differentiation of this function, the
initial rate of potential change was determined. The initial
potential change is a parameter independent of the permeability of
the lipid membrane. The permeability of the liposome membrane to
protons  is rather high (comparable to K$^{+}$ and Na$^{+}$ ions)
\cite{new:1990}. Therefore, estimation of the initial potential
change would be helpful in investigations of the electrogenic
activity of cytochrome-c oxidase. The further improvement of this
method in order to reduce the "stirring time" will allow the
recording of rapid kinetics of the membrane potential development
and, consequently, the resolving of proton translocation events
induced by cytochrome-c oxidase.
\end{section}
\clearpage

\begin{minipage}[t]{\textwidth}
\textit{Acknowledgements} \vspace{1cm}

\small The author wishes to thank Prof.~B.~Kadenbach,
Dr.~H.-J.~Apell and Dr.~R.J.~Clarke for many valuable discussions,
correspondence and suggestions concerning this work,
Dr.~E.~M{\"o}rschel for electronmicroscopic experiments and also
Mrs.~Sabine~Finkenstein for excellent technical assistance. This
work was financially supported by the Deutsche
Forschungsgemeinschaft (Ka 192/28-1).
\end{minipage}
\clearpage

\bibliography{Lit}
\end{document}